\documentclass[a4paper,twoside]{article}
\usepackage{ductri}
\usepackage{makecell}
\usepackage{enumitem}

\usepackage{epsfig}
\usepackage{subcaption}
\usepackage{calc}
\usepackage{amssymb}
\usepackage{amstext}
\usepackage{amsmath}
\usepackage{amsthm}
\usepackage{multicol}
\usepackage{pslatex}
\usepackage{apalike}
\usepackage{algorithm2e}
\usepackage[bottom]{footmisc}
\usepackage{SCITEPRESS}     

\begin{document}

\title{Nonlinear Model Predictive Control for Uranium Extraction-Scrubbing Operation in Spent Nuclear Fuel Treatment Process}

\author{\authorname{Duc-Tri Vo\sup{1,2}\orcidAuthor{0009-0006-4366-5258}, Ionela Prodan\sup{1}\orcidAuthor{0000-0002-3522-5192}, Laurent Lefèvre\sup{1}\orcidAuthor{0000-0002-5496-5882}, Vincent Vanel\sup{2}\orcidAuthor{0000-0001-8849-2174}, Sylvain Costenoble\sup{2}\orcidAuthor{0000-0003-0916-1006}, Binh Dinh\sup{2}\orcidAuthor{0000-0003-3076-619X}}
\affiliation{\sup{1}Univ. Grenoble Alpes, Grenoble INP, LCIS, F-26000, Valence, France.}
\affiliation{\sup{2}CEA, DES, ISEC, DMRC, Univ Montpellier, Marcoule, France.}
\email{\{duc-tri.vo, ionela.prodan, laurent.lefevre\}@lcis.grenoble-inp.fr, \{vincent.vanel, sylvain.costenoble, binh.dinh\}@cea.fr}
}


\abstract{This paper addresses the particularities of the uranium extraction-scrubbing operation in a spent nuclear fuel treatment process (PUREX-Plutonium Uranium Refining by Extraction) through the use of set-point tracking MPC (Model Predictive Control). The presented controller uses the feed solution flow rate as the manipulated variable to control the saturation of the solvent at the extraction step. In addition, it guarantees not to loose uranium in the raffinates, and ensures equipment limitations during operation time. Simulation results show that the tracking NMPC effectively ensures accurate set point tracking and constraints guarantee.  As a result, the system can be driven to its optimal working condition, avoid and recover from constraint violations. The control performance was compared with PID and openloop controllers.}

\onecolumn \maketitle \normalsize \setcounter{footnote}{0} \vfill

\section{\uppercase{Introduction}}
\label{sec:intro}
\subsection{PUREX Introduction and Motivation}
\begin{figure}[ht]
    \centering
    \includegraphics[width=0.9\linewidth]{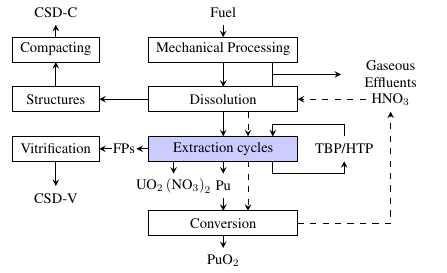}
    \caption{The PUREX process \cite{Vaudano2008}.}
    \label{fig:PUREX_main_steps}
\end{figure}

PUREX (Plutonium Uranium Refining by Extraction) is a hydro-metallurgical process used to recover and purify uranium and plutonium from spent nuclear fuels \cite{Vaudano2008}. It further allows the reuse of uranium and plutonium while ensuring that the nuclear waste is compatible with disposal requirements. The PUREX process is currently applied at an industrial scale at La Hague, a nuclear fuel reprocessing plant in northern France. As shown in Fig.~\ref{fig:PUREX_main_steps}, The PUREX process starts with removing the fuel cladding to permit nuclear material stored inside to be dissolved as wholly as possible in nitric acid. Next, fuel dissolution allows uranium and plutonium to be extracted and purified by liquid-liquid extraction techniques, which use tributyl phosphate (TBP) as an extractant in hydrogenated tetra propylene (HTP). Finally, uranium and plutonium are collected under the form of $\text{UO}_2\left(\text{NO}_3\right)_2$ and $\text{PuO}_2$ at the outlets of the process after conversion.

\begin{figure}[ht]
    \centering
    \includegraphics[width=\linewidth]{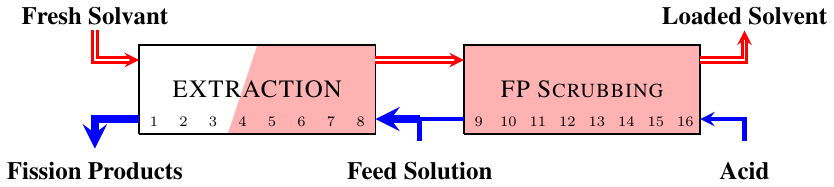}
    \caption{The extraction/FP scrubbing step.}
    \label{fig:system_simplified_EN}
\end{figure}

The primary control objective of the PUREX process is to quickly attain a high recovery rate and decontamination factor, disregarding the variations of system's parameters. A dedicated control strategy of the basic units in \textit{extraction cycles}, which are extraction-scrubbing, back extraction (stripping), and solvent generation  \cite{Dinh2008}, can be implemented in this aim. This research focuses on the uranium extraction-scrubbing process using mixer settlers, as depicted in Fig.~\ref{fig:system_simplified_EN}. Our model is described in details in Section~\ref{sec:process}. To the best of our knowledge, there are few similar works on control of this process in the literature. Consequently, we introduce in the next subsection available studies for similar processes, which are also liquid-liquid extraction using mixer settlers, as a good source of reference.

\subsection{Literature Review}

Among the first studies, the authors in \cite{Seemann1973}, modeled and developed a control scheme to maintain the efficiency of a rare earth elements extraction process under the variation of a measured system parameter. Based on the linearly approximated process model, this controller consists of three components: a pre-calculated control input at a steady state, a dynamic compensation term, and a proportional-integral (PI) correction term to eliminate the steady-state offset. Moreover, regarding the rare earth extraction operation, \cite{Yang2010}, \cite{Yang2016} used multiple linear models to identify the process and designed different model predictive controllers for each model. Then, the model-controller selection is made by considering the accumulative error during operation.

Another liquid-liquid extraction process using mixers-settlers is copper solvent extraction. In \cite{Komulainen2009} and \cite{Shahcheraghi2021}, a two-level optimization-stabilization control strategy was developed. At the optimization level, the optimal set-point that maximizes the process production is computed and fed to the stabilization level to stabilize the system at that set point. Multiple control schemes for the tracking layer, such as PI, PI combined with feed-forward control, Model Predictive Control, and Model Predictive Control combined with feed-forward control, were studied. The two-level control strategy reported in these works is popular in process control, as also discussed in \cite{Seborg2016}.

In the literature, there is no exactly similar process control problem as ours. However, we have seen that MPC was used in many applications. It is also a well-known control method in academia and industry \cite{Mayne2000}. In MPC, control inputs are obtained by solving an online constrained (nonlinear) optimization problem, with the current state as the initial condition. Consequently, MPC can efficiently optimize performance, handle constraints and nonlinearities, and ensure control stability. A comprehensive overview of MPC theory, computation, and implementation can be found in \cite{James2022}.

As previously mentioned, our particular process requires guarantees of hard constraints on process safety, performance, and equipment limits. Additionally, in the future, we want to exploit the advantage of the qualified simulation code PAREX \cite{Bisson2016} to compute the optimal control inputs online. Therefore, MPC is a promising approach for our control problem.

\subsection{Contributions and Paper Organization}
This paper introduces an optimal control technique for the uranium extraction-scrubbing process in the PUREX process, utilizing Nonlinear Model Predictive Control (NMPC) approach. The controller was designed to manipulate the feed solution flow rate while guaranteeing constraints which are not loosing uranium in the raffinates, and equipment limitations during operation time. Simulation results show that MPC effectively ensures accurate set point tracking and constraints guarantee. As a result, the system can be driven to its optimal working condition faster than the open loop case.

The paper is organized as follows. Section \ref{sec:process} introduces the process model, dynamic characteristics, set up of the control problem and the state-space representation. Then, the NMPC is formulated in Section \ref{sec:NMPC}. In Section \ref{sec:case_studies}, simulation results over different cases studies are presented. Finally, conclusions are stated in Section \ref{sec:conclusion}.

\textbf{Notations:} For $\bz \in \mathbb{R}^n$, we denote $\left\|\bz\right\|_\bQ^2 = \bz^T \bQ \bz$. $u(k-1|k):= u(k-1)$. $\bI$ denotes identity matrix of approriate dimension. Inequalities between vectors are evaluated element-wise.

\textbf{Remark:} For the sake of data confidentiality, some numerical parameters of the process are normalized. The nominal parameter values are denoted with a $0$ superscript.

\section{\uppercase{Uranium Extraction-Scrubbing}} \label{sec:process}
\begin{figure*}[ht!]
    \centering
    \includegraphics[width=\linewidth]{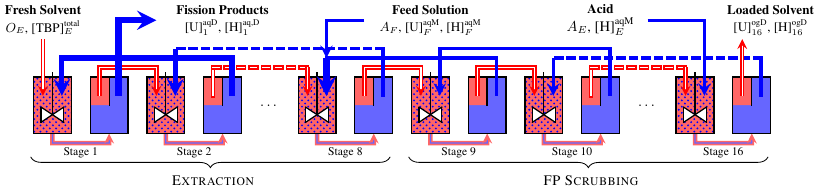}
    \caption{Uranium extraction-scrubbing operations using mixers-settlers}
    \label{fig:system_detailed_EN}
\end{figure*}

\subsection{Mathematical Model}
The principle of the extractor is depicted in Fig.~\ref{fig:system_detailed_EN}. It consists of 16 stages of mixer-settlers in series, which is typical of the uranium extraction-scrubbing process tests achieved at CEA. Our mathematical model is based on mass balance equations and the assumptions: (i) constant density, (ii) immiscibility of the aqueous and organic phases, (iii) perfect mixing in the mixer, (iv) transfer kinetics is disregarded. System parameters notation are described in Fig.~\ref{fig:system_detailed_EN} and Tab.~\ref{tab:system_params}. Equation \eqref{eq:equilibrium} describes the primary extraction mechanism.
\begin{subequations}
\begin{align}
    \text{UO}_2^{2+} + 2\text{NO}_3^{-} + 2\text{TBP} &\overset{K_U}{\rightleftharpoons} \text{UO}_2\left(\text{NO}_3\right)_2\cdot \text{TBP}\label{eq:equi_eqs1}\\
    \text{H}^+ + \text{NO}_3^{-} + \text{TBP} &\overset{K_H}{\rightleftharpoons} \text{HNO}_3\cdot \text{TBP}\label{eq:equi_eqs2}
\end{align}\label{eq:equilibrium}
\end{subequations}

\begin{table}[h] 
    \centering
    \caption{System parameters notation and descriptions.}
    \begin{tabular}{ll}
    \hline
    \textbf{Notation} & \textbf{Description} \\ \hline
    $A$, $O$ & Aqueous and organic flow rates.\\
    $V$, $W$ & Aqueous and organic volumes.\\
    $\KU$, $\KH$ & Equilibrium constants for $U$ and $H$.\\
    $k_U$, $k_H$ & Mass transfer coefficients for $U$, $H$.\\
    $^\text{aq}$, $^\text{og}$ & Related to aqueous and organic phase.\\
    $^\text{M}$, $^\text{D}$ & Related to mixer and settler.\\
    $[\cdot]$ & Concentration.\\
    $_n$ & Related to stage $n$ of the process.\\
    $_i$ & Related to inputs to stage $n$. \\ \hline    
    \end{tabular}
    \label{tab:system_params}
\end{table}

\begin{figure}[ht]
    \centering
    \includegraphics[width=0.8\linewidth]{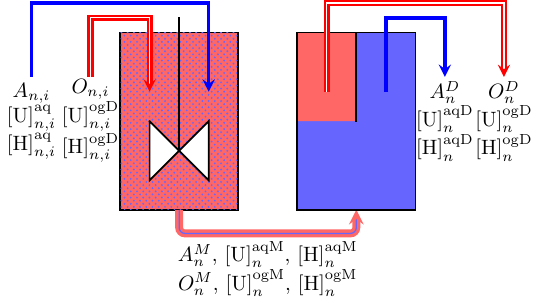}
    \caption{Mixer-settler model.}
    \label{fig:SysDiagram_2}
\end{figure}

Assuming that mixer volumes satisfy:
\begin{subequations}
\begin{gather*}
    \dot{V}^M_n = \dot{W}^M_n = 0 \Rightarrow \dfrac{A_{n,i}}{O_{n,i}} = \dfrac{\VM{n}}{\WM{n}}\cdot
\end{gather*}
\end{subequations}

Uranium mass balances in mixers and settlers are given in \eqref{eq:mass_balance_U}. We assume that the mass transfer term of uranium from the organic to aqueous phase is given by $\Phi_n^U$, where the mass transfer coefficient $k_U$ is very large. $\UogM{n,*}$ is computed from the chemical equilibrium condition.

\begin{subequations}
\begin{align}
    \VM{n} \dUaqM{n} &= \An{n,i}\text{[U]}^\text{aq}_{n,i} - A^M_n \UaqM{n} + \Phi_n^U\\
    \WM{n} \dUogM{n} &= \On{n,i}\text{[U]}^\text{og}_{n,i} - O^M_n \UogM{n} - \Phi_n^U\\
    \VD{n} \dUaqD{n} &= A^M_n \UaqM{n} - A^D_n\UaqD{n}\\
    \WD{n} \dUogD{n} &= O^M_n \UogM{n} - O^D_n\UogD{n}\\
    \Phi_n^U &= k_U\left(\UogM{n,*} - \UogM{n}\right)\\
    \UogM{n*} &= \KU \UaqM{n}\NO{n}^2\TBPfree{n}^2 \label{eq:KU}
\end{align}\label{eq:mass_balance_U}
\end{subequations}

Mass balances equations for $H^+$ can be obtained by replacing $U$ by $H$ in \eqref{eq:mass_balance_U}, note that $\HogM{n,*}$ is computed as follow:
\begin{gather}
    \HogM{n*} = \KH \HaqM{n} \NO{n} \TBPfree{n}
\end{gather}
Additionally, we have equations for total aqueous nitrate ion concentration and total organic TBP concentration in mixer \eqref{eq:TBP_NO}.
\begin{subequations}
\begin{gather}
    \NO{n} = 2\UaqM{n} + \HaqM{n} \label{eq:NO3}\\
    \TBPtotal{n} = \TBPfree{n} + 2 \UogM{n} + \HogM{n}.\label{eq:TBPfree}
\end{gather}\label{eq:TBP_NO}
\end{subequations}
We also assume that
\begin{subequations}
\begin{gather}
    A_n := A_n^M = A_n^D,\quad O_n := O_n^M = O_n^D,\\
    \VM{1} + \WM{1} = \VM{2} + \WM{2} = \dots = \VM{16} + \WM{16}\\
    \TBPtotal{E} = \TBPtotal{1} = \dots = \TBPtotal{16}. 
\end{gather}\label{eq:addtional_eqs}
\end{subequations}

\textbf{Remarks:} When $k_U, k_H$ is large, the system dynamics becomes stiff. Additionally, the system dynamics is high dimensional with 128 states. Hence, the MPC optimization problem for this problem is an large-scale Nonlinear Programming Problem (NLP) that requires high computational efforts. As a consequence, we want to study the MPC behavior with a similar and simpler dynamics. Therefore, in this first study, to facilitate numerical computations, we assume that $\dUogM{n}= 0$. Note that this assumptions should be removed in future studies.

\subsection{Control Objectives}
Since this is the first study on control of this process using MPC, for the work in this manuscript, we chose the control objective as follow. We aim to maximize the amount of extracted uranium \eqref{eq:R}, while keeping uranium concentration in fission products below a tolerance \eqref{eq:cons_fs} and constraints on control inputs. The final time $t_f$ in \eqref{eq:R} can be chosen as the settling time of the open loop system. In future works, we will aim to control the solvent saturation level.
\begin{subequations}
\begin{gather}
    R = \int_{0}^{t_f} O^D_{16} \UogD{16} \text{d}t \label{eq:R}\\
    \UaqD{1} \le \UaqD{1,\text{tol}} \label{eq:cons_fs}\\
    A_{F,\text{min}} \le A_F \le A_{F,\text{max}}\\
    \Delta A_{F,\text{min}} \le \Delta A_F \le \Delta A_{F,\text{max}}
\end{gather}
\end{subequations}

\subsection{Solvent Saturation and Set Point Determination}\label{sec:SSC}

\begin{figure}[ht]
    \centering
    \includegraphics[width=0.9\linewidth]{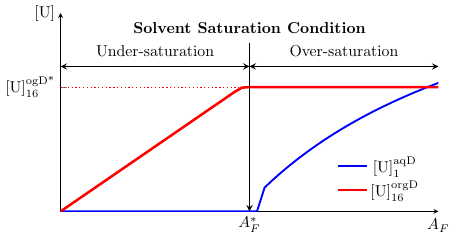}
    \caption{Steady state relationship of feed solution flow rates and uranium concentrations.}
    \label{fig:SSC}
\end{figure}

Analyzing the process dynamics shows a critical operating condition in which the system becomes saturated. As an illustration, Fig.~\ref{fig:SSC} shows the steady state relationship between the feed solution flow rate $A_F$ and uranium concentrations at the system outlets.

It can be seen from Fig.~\ref{fig:SSC} that if the solvent is under-saturated, by increasing $A_F$, we can increase the extracted uranium until the solvent saturation condition is reached $\left(A_F = A_F^*\right)$. However, once the solvent is over-saturated, increasing $A_F$ no longer increases the amount of uranium extracted but drastically increases $\UaqD{1}$. This behavior is undesirable because a large amount of uranium will be entrained in raffinates with the non-recyclable species, such as fission products. Since $A_F^*$ has a very small margin towards the over-saturation zone, if we succeed in controlling the system to work at $A_F^*$, we can manage in other under-saturated set points. Note that in practice, the set point should be chosen appropriately depending on the equipment’s limits.

\subsection{Manipulated, Disturbance and Controlled Variables}
Since this system dynamic is nonlinear and high dimensional with 96 stages, it is preferable to begin with a control scheme with one control input and one disturbance variable, $A_F$ and $O_E$ respectively. This choice is based on the fact that the flow rate is much easier to manipulate than the concentration. It can be done by using commercial pump controllers and flow meters. In addition, by observing system dynamics, $A_F$ and $O_E$ are much more sensitive to uranium concentrations compared to $A_E$. It can be verified that this choice makes the system stabilizable and output controllable.

Since MPC has the advantage in multi-variable control, a state feedback control approach is chosen. On the contrary, for single input single output controller such as PID, $\UogD{16}$ is chosen as the controlled variable. It should be noted that although sensors have limitations to measure uranium concentrations in organic phase, we can still obtain this value from the mathematical model. States estimation, noises and model mismatches are not covered in this manuscript, all the states are assumed to be well feed-backed.

\subsection{State Space Representations}
The process dynamics described in \ref{sec:process} can be written as a system nonlinear ordinary differential equations (ODEs). The continuous state space representation of the system can be represented as follow:
\begin{gather}
    \bdx = \ff_c(\bx,\, u,\, p),\label{eq:state_space}
\end{gather}
in which:
\begin{itemize}
    \item $ \bx = \tmat{\bx_{\left(1\right)}^T & \bx_{\left(2\right)}^T & \bx_{\left(3\right)}^T & \bx_{\left(4\right)}^T & \bx_{\left(5\right)}^T & \bx_{\left(6\right)}^T}^T$ are system states, $ \bx \in \mathbb{R}_{96\times 1}$, in which
    \begin{small}
        \begin{align*}
        \bx_{\left(1\right)}^T&=\tmat{\UaqM{1} & \UaqM{2} & \dots & \UaqM{16}},\\
        \bx_{\left(2\right)}^T&=\tmat{\UaqD{1} & \UaqD{2} & \dots & \UaqD{16}},\\
        \bx_{\left(3\right)}^T&=\tmat{\UogD{1} & \UogD{2} & \dots & \UogD{16}},\\
        \bx_{\left(4\right)}^T&=\tmat{\HaqM{1} & \HaqM{2} & \dots & \HaqM{16}},\\
        \bx_{\left(5\right)}^T&=\tmat{\HaqD{1} & \HaqD{2} & \dots & \HaqD{16}},
        \end{align*}
        
        \begin{align*}
        \bx_{\left(6\right)}^T&=\tmat{\HogD{1} & \HogD{2} & \dots & \HogD{16}};
        \end{align*}
    \end{small}
    \item $u = \An{F}$ is the manipulated variable;
    \item $p = \On{E}$ is the measured varying parameter;
    \item $\ff_c$ is the vector of mass balance equations of corresponding states in $\bx$.
\end{itemize}

Denoting $\bbx\left(\Bar{k}\right) := \bx(\Bar{k}h)$, the discrete time model can be obtained by using the Euler method with sampling time $h\in \mathbb{R}^+$ and $\Bar{k} \in \mathbb{Z}^+$  as follows:
\begin{align*}
    \bbx(\Bar{k}+1) &= \bbx\left(\Bar{k}\right) + h \ff_c (\bbx\left(\Bar{k}\right),\, u\left(\Bar{k}\right),\, p\left(\Bar{k}\right)).
\end{align*}
From our experience, to ensure the discretization convergence, $h$ should be sifficiently small, about $10^{-3}$ hours. However, since the process dynamic is slow, the control sampling time $T_s$ is much larger, about 0.1 or 0.5 hours for example. Assume that $u(t), \, p(t)$ are constant $\forall t\in \left[kT_s, (k+1)T_s\right)$, let $N_s := T_s/h$, it is more convenient to denote $\bx(k) := \bx\left(kT_s\right)$ and
\begin{align}
    \bx(k+1) &=  \ff(\bx(k),u(k),p(k)).\label{eq:discrete_state_space}
\end{align}

\section{\uppercase{NMPC Formulation}} \label{sec:NMPC}
In this section, we formulate the set-point tracking MPC. In general, it has two steps:
\begin{enumerate}[label=(\roman*)]
    \item defining the appropriate set point depending on the value of $p$ in \eqref{eq:state_space} and the flow sheet;
    \item apply the MPC or PID controller to stabilize the system at the defined set point.
\end{enumerate}

Consider the discrete model \eqref{eq:discrete_state_space}, given the initial state $\bx(k)$, denote $\bx(j|k)$ the predicted state at time step $j$ driven by the predicted control inputs $\{u(i|k)\}_{i=k}^{j}$, $\forall j \in \mathbb{Z}^+$, $j>k$. In addition, denote $\left(\bx_\text{set}, u_\text{set}\right)$ the desired set point and $\btx := \bx - \bx_\text{set}$, $\tilde{u}:=u - u_\text{set}$, the quadratic cost function can be defined as follows, with $N_p$ denotes the prediction horizon:
\begin{align*}
    &\ell \left(\bx(k), u(k-1), \{u(i|k)\}_{i=k}^{N_p-1}\right) \notag\\
    =&\sum_{i=k}^{N_p-1}\left(\left\|\btx(i|k)\right\|_{\bQ}^2 + \left\|\tilde{u}(i|k)\right\|_{\bR}^2 \right) \notag\\
    &+  \sum_{i=k}^{N_p-1} \left\|u(i|k) - u(i-1|k)\right\|_{\bS}^2 + \left\|\btx(N+1|k)\right\|_{\bP}^2 
\end{align*}
in which $\bQ,\, \bR, \bP$, $\bS$ denote symmetric positive definite weighting matrices. The constraints \eqref{eq:constraints} can be written as follow, for all $i,\, k \in \mathbb{Z}^+,\; i\in \left[k, k+N_p\right]$:
\begin{subequations}
\begin{gather}
    \bx(i|k) \ge 0\\
    \bx_{17} (i|k) \le \UaqD{1,\text{tol}}\\
    A_{F,\text{min}} \le u(i|k) \le A_{F,\text{max}}\\
    \Delta A_{F,\text{min}} \le u(i|k) - u(i|k-1) \le \Delta A_{F,\text{max}}.
\end{gather}\label{eq:constraints}
\end{subequations}

In summary, at each control sampling time $k$, given the state vector $\bx(k)$, MPC solves the following optimization problem to obtain optimal open loop control inputs $\{u^*(i|k)\}_{i=k}^{N_p-1}$. Then, the first control input $u^*(k|k)$ is applied to the system until the next time step $k+1$, hence closing the loop.
\begin{align*}
\min_{\{u(i|k)\}_{i=k}^{N_p}} \ell \left(\bx(k), u(k-1), \{u(i|k)\}_{i=k}^{N_p-1}\right)
\end{align*}
subject to \eqref{eq:constraints}, $\forall i,\, k \in \mathbb{Z}^+,\; i\in \left[k, k+N_p - 1\right],$ and:
\begin{align*}
    \bx(i+1|k) &= \ff(\bx(i|k),u(i|k),p(i|k)),\\
    \bx(k|k) &= \bx (k).
\end{align*}

\section{\uppercase{Case Studies}}\label{sec:case_studies}
This section presents the simulation studies of the nonlinear tracking MPC approach for the system shown in Fig.~\ref{fig:system_detailed_EN}. Set point tracking with varying parameters and constraint-handling applications are studied. NMPC performance is compared to open loop and PID controllers. The NMPC implementation was based on CasADi toolbox \cite{Andersson2019} and the nonlinear programming (NLP) solver IPOPT \cite{Wachter2005c}. In addition, SUNDIALS, cf. \cite{hindmarsh2005sundials} and \cite{gardner2022sundials}, was also used as differential and algebraic equation solvers.

\begin{figure}[ht]
    \centering
    \includegraphics[width=\linewidth]{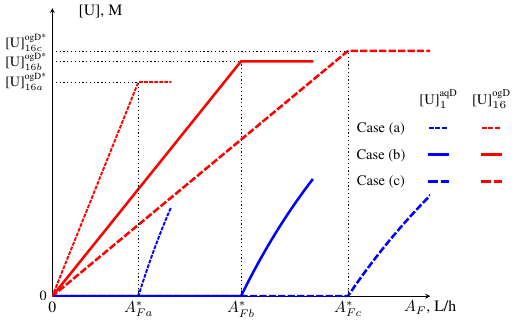}
    \caption{Steady state relationship of $A_F$ and $\UogD{16}$ in 3 cases (a): $O_E = 0.5O_E^0$, (b) $O_E = O_E^0$, and (c) $O_E = 1.5 O_E^0$.}
    \label{fig:SSC_3cases}
\end{figure}

The MPC weighting matrices were chosen heuristically, with $\bQ=\bP=\bI$ and $\bR=\bS=1/u_\text{set}$. The control sampling time $T_s$ is 0.5 hours and the prediction horizon $N_p$ is 10 steps. In addition, set points are chosen as saturation points, as shown in Fig.~\ref{fig:SSC_3cases}. Note that \textit{critical set points} must be redefined whenever $O_E$ changes to ensure the feasibility and optimality of the control objectives. Finally, $\UaqD{1}$ is plotted in logarithmic scale.

The PID controller with $y=\UogD{16}$ is given as:
\begin{subequations}
\begin{align}
    u(k) &= u_\text{set} + K_P e(k) + K_I e_I(k) + K_D e_D(k)\\
    e(k) &= y_\text{set}-y(k)\\
    e_I(k) &= 0.5 T \left[e(k) + e(k-1) \right] + e_I(k-1)\\
    e_D(k) &= \left[e(k) - e(k-1)\right]/T
\end{align}\label{eq:PID}
\end{subequations}
The gains $K_P,\, K_I, \, K_D$ are tuned by solving the following optimization offline:
\begin{align*}
    \min_{K_P,\, K_I, \, K_D} \sum_{k=0}^{N_\text{PID}-1}  \Big[ e^2(k) + r\tilde{u}^2 (k)+ s\left[u(k+1)-u(k)\right]^2\Big]
\end{align*}
subject to \eqref{eq:discrete_state_space}-\eqref{eq:PID} with $r$, $p$, $s$ are weighting parameters. $N_\text{PID}$ was chosen to be 30 steps, which is the settling time of the open loop system. Finally, saturation is also added to satisfy bounded constraints on control inputs, $\forall k \in \mathbb{Z}^+$:
\begin{gather*}
    u(k) = \left\{ \begin{aligned}  &u_\text{min},&  u\le u_\text{min}\\
     &u_\text{max},& u \ge u_\text{max}\\
      &u(k),&  \text{otherwise.}
    \end{aligned} \right.\\
    u(k) = \left\{ \begin{aligned}  &u(k-1) + \Delta u_\text{min},&  u(k)-u(k-1)\le \Delta u_\text{min}\\
    &u(k-1) + \Delta u_\text{max},& u(k) - u(k-1) \ge \Delta u_\text{max}\\
    &u(k),&  \text{otherwise.}\\
    \end{aligned} \right.
\end{gather*}

\subsection{Nominal Set Point Tracking with Varying Parameters}
\begin{figure}[ht]
    \centering
    \includegraphics[width=\linewidth]{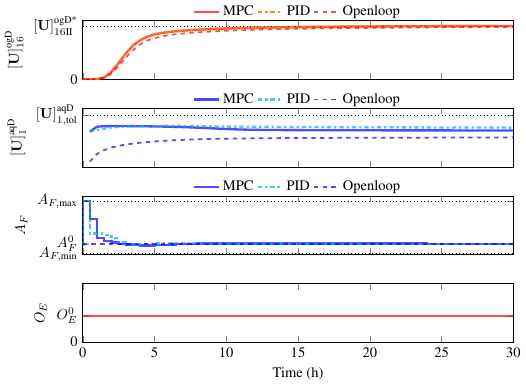}
    \caption{Set-point tracking application with MPC, PID and open loop controllers.}
    \label{fig:setpoint_tracking}
\end{figure}

In this subsection, it is assumed that $t=0$, the system is at steady state with nominal parameters except for $\UaqF = 0$. In other words, uranium is only sent into the system once $t>0$.  The simulation results for the \textit{nominal case} is presented in Fig.~\ref{fig:setpoint_tracking}. It is shown that both NMPC, PID, and open loop controllers can stabilize the output at the set point while adhering to the constraint on $\UaqD{1}$. A quantitatively comparison can be done by computing the amount of extracted uranium $R$ given in \eqref{eq:R}. Since the 2\% settling time of the open loop case is 30 hours, approximate \eqref{eq:R} by \eqref{eq:R_approx}, and choose $t_f = 30$ hours (thus $k_f = 60$ steps). Finally, the amount of extracted uranium for MPC and PID is 4\% higher than the open loop case.
\begin{align}
    R \approx T_s \sum_{k=0}^{k_f} O_E(k) \UogD{16} (k)\label{eq:R_approx}
\end{align}

\begin{figure}[ht]
    \centering
    \includegraphics[width=\linewidth]{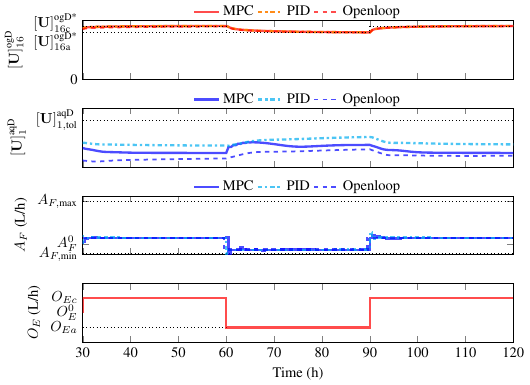}
    \caption{Set-point tracking application subject to disturbances with MPC, PID and open loop controllers.}
    \label{fig:setpoint_tracking_2}
\end{figure}

If we continue the simulation in Fig.~\ref{fig:setpoint_tracking} and vary $O_E$ as a varying parameter, the simulation result is shown in Fig.~\ref{fig:setpoint_tracking_2}. The disturbances appear at 30h, 60h, and 90h with $O_E$ increase or decrease of 50\% of $O_E$ from its nominal value, which is an important point in the process. Consequently, the system needs to be stabilized at its new optimal operating points shown in Fig.~\ref{fig:SSC_3cases}. Furthermore, the differences are not so significant, they converge to the openloop case.

\subsection{Constraint Guaranteeing}
\begin{figure}[ht]
    \centering
    \includegraphics[width=\linewidth]{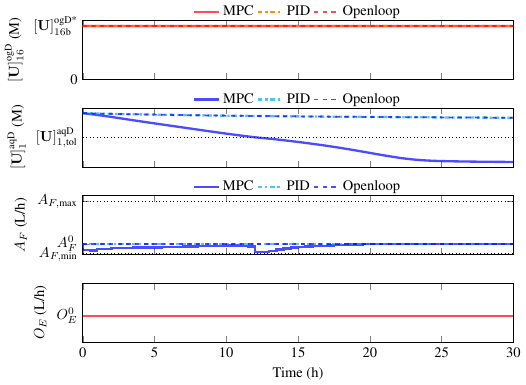}
    \caption{Set-point tracking with infeasible initial condition with MPC, PID, and open loop controllers.}
    \label{fig:setpoint_tracking_3}
\end{figure}

One interesting question is that if, due to some reasons, the system is over-saturated, can the controllers eliminate the constraint violation and stabilize the system at its set point? Fig.~\ref{fig:setpoint_tracking_3} illustrates this scenario. The system is assumed to be initially steady and over-saturated in this simulation. It can be seen that only NMPC can quickly reduce $\UaqD{1}$ to its tolerance. This result can be explained by the fact that MPC considers all the state errors while PID only focuses on one controlled variable, $\UogD{16}$. In addition, NMPC explicitly handles the constraint \eqref{eq:constraints} at every time step. Furthermore, since $\UaqD{1}$ is very sensitive to $A_F$, although the steady state control inputs are similar (but not exactly equal), we have seen a significant different in $\UaqD{1}.$

\section{\uppercase{Conclusion}}\label{sec:conclusion}
This paper presents an NMPC (Nonlinear Model Predictive Control) approach for the uranium extraction-scrubbing operation in the PUREX process. It was shown that this approach favors the process control objectives in stabilizing the system at the optimal working condition with constraints satisfaction. As a result, the process performance was increased quantitatively in terms of the amount of extracted uranium. This study provides a good reference for future developments on controlling extraction cycles in the PUREX process. Constraint handling is the key factor that makes MPC more beneficial for practical applications than the classical PID. Future developments include stability guarantees, uncertainties handling, and verification with the qualified simulation code PAREX \cite{Bisson2016} as a virtual plant in multiple application scenarios. Moreover, future studies will be conducted at more sensitive point in the process. Furthermore, the development of an observer is essential to provide an output feedback MPC scheme with limited measurements. Finally, experiments will be conducted to evaluate the practical implementation aspects of the developed control scheme.

\section*{ACKNOWLEDGEMENTS}
The authors thank ORANO for partial financial support for the project.

\bibliographystyle{apalike}

\end{document}